\documentclass[8.5pt,twoside,twocolumn]{article}
\usepackage[super,sort&compress,comma]{natbib} 
\usepackage{mhchem}
\usepackage{times}
\usepackage{sectsty}
\usepackage{balance} 

\usepackage{graphicx} 
\usepackage{lastpage}
\usepackage[format=plain,justification=raggedright,singlelinecheck=false,font=small,labelfont=bf,labelsep=space]{caption} 

\begin{document}

\makeatletter 
\def\subsubsection{\@startsection{subsubsection}{3}{10pt}{-1.25ex plus -1ex minus -.1ex}{0ex plus 0ex}{\normalsize\bf}} 
\def\paragraph{\@startsection{paragraph}{4}{10pt}{-1.25ex plus -1ex minus -.1ex}{0ex plus 0ex}{\normalsize\textit}} 
\renewcommand\@biblabel[1]{#1}            
\renewcommand\@makefntext[1]%
{\noindent\makebox[0pt][r]{\@thefnmark\,}#1}
\makeatother 
\renewcommand{\figurename}{\small{Fig.}~}
\sectionfont{\large}
\subsectionfont{\normalsize} 

\twocolumn[
  \begin{@twocolumnfalse}
\noindent\LARGE{\textbf{Templated self-assembly of patchy particles}}
\vspace{0.6cm}

\noindent\large{\textbf{Alexander J Williamson,\textit{$^a$} Alex W Wilber,\textit{$^a$} Jonathan P K Doye,$^{\ast}$\textit{$^a$} and Ard A Louis\textit{$^b$}}}\vspace{0.5cm}

\vspace{0.6cm}

\noindent \normalsize{We explore the use of templated self-assembly to facilitate the formation
of complex target structures made from patchy particles.
First, we consider the templating of high-symmetry shell structures around 
a spherical core particle. We find that nucleation around the core particle
can inhibit aggregate formation, a process which often hinders self-assembly.

In particular, this new assembly pathway allows dodecahedral shells to 
form readily, whereas these structures never form in the absence of the template.
Secondly, we consider the self-assembly of multi-shell structures, where the
central icosahedral core is known to form readily on its own, and which could
then template the growth of further layers. We are able to find conditions 
under which two- and three-shell structures successfully assemble, illustrating
the power of the templating approach.}
\vspace{0.5cm}
 \end{@twocolumnfalse}
  ]


\section{Introduction}


\footnotetext{\textit{$^{a}$~Physical and Theoretical Chemistry Laboratory,
  Department of Chemistry, University of Oxford, South Parks Road,
  Oxford, OX1 3QZ, United Kingdom}}
\footnotetext{\textit{$^{b}$~Rudolf Peierls Centre for Theoretical Physics,
 University of Oxford, 1 Keble Road, Oxford, OX1 3NP, United Kingdom}}



It has been a long-held goal to use self-assembly to create complex, ordered 
structures on the micro- and nanoscale.\cite{Whitesides02b}
Biological systems with their dazzling array of ordered and precise 
self-assembled structures \cite{Goodsell04} show what can be possible 
given sufficient control of the design and the interactions between
the building blocks. Thus, with the goal of taking the first steps to 
achieving similar control in synthetic systems, researchers working on 
nanoparticles and colloids are making great efforts to synthesize 
``patchy'' particles that have interactions only in specific 
directions.\cite{Glotzer07,Roh06,Cho07,Devries07,Wang08,Perro09,Kraft09,Pawar10}

The formation of icosahedral virus capsids, proteinaceous shells with specific 
size and structure that are designed to encapsulate the viral genome, 
provides one of the archetypal examples of biological self-assembly, 
and also one of the most studied.\cite{Zlotnick03b}
If similar structures are to be achieved in synthetic systems, it will
be important to understand the basic physical principles of such self-assembly
and the design rules for the interactions between the constituent particles.
So far, theory \cite{Endres02,Endres05,Zlotnick05,Zandi06} and 
simulations \cite{Hagan06,Nguyen07,Nguyen08,Nguyen08b,Zhang06b,Sweeney08,Rapaport04,Rapaport08,Wilber09b,Johnston10,Wales05,Fejer09,Fejer10} have been most concerned 
with understanding the assembly of smaller ($T=1$, 3 or 4) empty capsids. 
However, functional viruses are not empty, but contain the genomic material. 
For icosahedral RNA viruses, the capsid 
proteins and the RNA typically co-assemble, and experiments have indicated
that the kinetics of such assembly can be different from that
of empty capsids.\cite{Johnson03} Indeed, this ability of capsid
proteins to encapsulate is not limited to the viral genome, 
but has been exploited to achieve capsid assembly around 
nanoparticles,\cite{Dragnea03,Chen06,Sun07} 
nanoemulsion droplets \cite{Chang08} and anionic polymers.\cite{Hu08}
Furthermore, for larger and more complex capsids, 
assembly can be dependent on the presence of ``scaffolding'' proteins, 
which are thought to template the correct assembly of the 
capsid.\cite{Maraver03,Deng08}

Although less studied, simulations \cite{Hagan08,Elrad08,Kivenson10,Elrad10}
and theory \cite{vanderSchoot05,Hu07,Hagan09} have begun to address the
assembly of filled capsids. Particularly relevant to the present study are
the simulations of Hagan and coworkers, who have shown that the change in 
assembly mechanism from homogeneous nucleation of the capsid proteins in the 
empty capsid case to heterogeneous nucleation around a core enhances the ability
of the capsids to assemble.\cite{Hagan08}

When considering the design of synthetic systems of patchy particles which assemble into capsid-like structures, one has to consider the potential differences in 
the interactions between the biological and synthetic cases. For capsids, 
the interfaces responsible for the protein-protein interactions
not only have to be in contact, but have to have the correct relative orientation.
However, first-generation patchy colloids and nanoparticles are unlikely to 
have this ``torsional'' component in the potential. Simulations have indicated
that the one-component self-assembly of simple monodisperse targets is 
still feasible in the absence of torsional constraints;\cite{Zhang04,Wilber07,Wilber09} however, there can be significant differences in the mechanisms of assembly and the nature of the kinetic 
traps compared to virus capsids.  In particular, the lack of a torsional 
component in the interparticle potential to enforce convexity in the growing 
clusters leads to disordered aggregation competing with correct 
assembly.\cite{Wilber07,Wilber09} Furthermore, as 
the size of the target structure increases, the difficulty of assembly 
increases much more rapidly when torsional constraints are not 
present. For example, the self-assembly of 20-particle dodecahedra occurs 
readily with a protein-like potential with torsional constraints,\cite{Wilber09b}  
but is seemingly impossible without.\cite{Wilber09}

Here, we will explore whether templated assembly might potentially 
provide a way to allow the formation of more complex targets using synthetic
patchy particles. We will consider both assembly around a single central core 
particle (Section \ref{sect:core}) and also the assembly of two- and 
three-shell structures (Section \ref{sect:multi}), where the central shell is 
known to readily assemble on its own.\cite{Wilber07}
If the materials for the different particles were chosen appropriately, 
the template could then be selectively removed 
by chemical\cite{Vlasov01} or thermal\cite{Velikov02} treatment, if desired.

We should also note that templating has been a useful strategy in supramolecular chemistry,\cite{Miras10} and even for the macroscopic assembly of magnetic subunits into polyhedral shells.\cite{Burnley05}






\section{Methods}
\label{sect:methods}
\subsection{Potential}
To model the patchy particles, we use the potential that we have used in 
our previous work on self-assembly,\cite{Wilber07,Wilber09} but with the 
additional feature that we consider multiple particle types.  
The model has also been used to study the crystallization of patchy colloids, \cite{Doye07,Noya07b,Noya10} and, with an additional torsional component to the 
potential, the self-assembly of protein complexes.\cite{Wilber09b,Villar09}

In the model, the repulsion between particles $i$ and $j$ is based upon an 
isotropic Lennard-Jones potential
\begin{equation}
V_{\rm LJ}(r_{ij}) = 4\varepsilon_{\rm ref}\left[ \left( \frac{\sigma_{ij}}{r_{ij}}
    \right)^{12} - \left( \frac{\sigma_{ij}}{r_{ij}} \right)^{6} \right], 
\label{eq:LJ} 
\end{equation}
but where the attraction is modulated by an orientational term,
$V_{\rm ang}$.  Thus, the complete potential is
\begin{equation}
V({\mathbf r_{ij}},{\mathbf \Omega_i},{\mathbf \Omega_j})=\left\{
    \begin{array}{ll}
       V_{\rm LJ}(r_{ij}) & r_{ij}<\sigma_{ij} \\
       V_{\rm LJ}(r_{ij})
       V_{\rm ang}({\hat{\mathbf r}}_{ij},{\mathbf \Omega_i},{\mathbf \Omega_j})
                       & r_{ij}\ge \sigma_{ij}, \end{array} \right.
\label{eq:LJaniso} 
\end{equation}
where ${\mathbf \Omega_i}$ is the orientation of particle $i$, and
\begin{equation}
\begin{split}
V_{\rm ang}
({\mathbf {\hat r}_{ij}},{\mathbf \Omega_i},{\mathbf \Omega_j})
=\max\left[\frac{\varepsilon_{\alpha\beta}}{\varepsilon_{\rm ref}} 
\exp\left(-{\theta_{\alpha ij}^2\over 2\sigma_{{\rm pw},\,\alpha}^2}\right)\right.\\ 
\left. \times \exp\left(-{\theta_{\beta ji}^2\over 2\sigma_{{\rm pw},\,\beta}^2}\right) 
\right]
\end{split}
\end{equation}
where 
$\theta_{\alpha ij}$ is the angle between the normal to patch $\alpha$ 
on particle $i$ and the interparticle vector $\mathbf r_{ij}$,
and the `$\max$' selects 
the pair of patches that have the strongest interaction for the current 
geometry. 
We assume that the particle sizes are additive, 
\emph{i.e.}\ $\sigma_{ij}=(\sigma_{ii}+ \sigma_{jj})/2$.
We also generally choose $\sigma_{\rm pw}$, a measure of the width of 
a patch, to be the same for all patches, except in the case of the central
particle in Section \ref{sect:core}.
By contrast, we allow the well-depth of the patch-patch interactions, 
$\varepsilon_{\alpha\beta}$ to 
vary ($\varepsilon_{\rm ref}=\max[\varepsilon_{\alpha \beta}]$).
In addition, for computational efficiency, we cut and shift the potential
at $3.5\,\sigma_{\text{max}}$ 
(where $\sigma_{\text{max}}$ is $\sigma$ for the largest particle in the system), and also shift the crossover distance in 
Eq.\ (\ref{eq:LJaniso}) so that it still occurs when the potential is zero.

We note that in this model, we have effectively coarse-grained out the solvent
degrees of freedom, and so when we talk about a gas phase in our model, this
would correspond to a dilute solution.

\subsection{Simulations}
We use the virtual move Monte Carlo algorithm of Whitelam and Geissler 
\cite{Whitelam07,Whitelam08} to simulate the dynamics. We choose this algorithm
for a number of reasons. Firstly, as with other Monte Carlo algorithms based on
local moves, the algorithm gives diffusive dynamics, as is appropriate for a 
model of colloids and nanoparticles in solution. Secondly, it can generate 
collective motion of the particles. Thirdly, the algorithm is designed to 
generate the correct relative diffusion rates for clusters of different size. 
The latter two features represent potential advantages over the 
single-particle Monte Carlo we typically used in our previous studies.\cite{Wilber07,Wilber09,Wilber09b}
 However, as the main mode of cluster growth
in those studies was by monomer addition, this was not a significant drawback. 
Here, however, we felt that it was important not to artificially disfavour 
growth by the addition of clusters particularly in the case of the 
multi-shell-structures studied in Section \ref{sect:multi}. As the actual
algorithm is quite involved, we do not give the details here, but refer the 
interested reader to the original papers.\cite{Whitelam07,Whitelam08} 
We note that we use a maximum translational move size of $0.3\,\sigma_{\text{min}}$ 
(where $\sigma_{\text{min}}$ is $\sigma$ for the smallest particle in the system), as we 
have found this to give a good balance between moves that lead to internal
rearrangements within clusters and moves that lead to diffusive behaviour of
complete clusters.\cite{WilberThesis2009}

\begin{figure}
\centering
\includegraphics{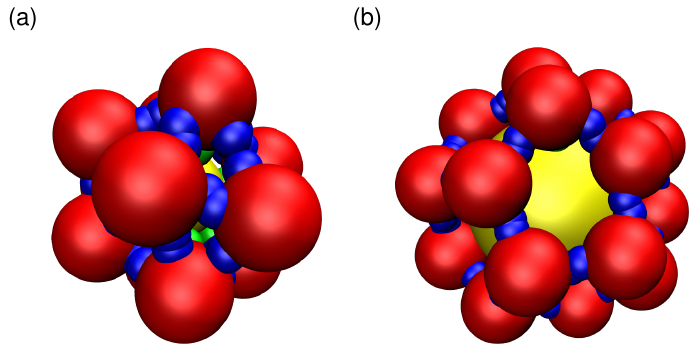}
\caption{\label{fig:targets} Target geometries: (a) AB$_{12}$, 
a centred icosahedron and (b) AB$_{20}$, a centred dodecahedron.
}
\end{figure}

\section{Assembly around a core particle}
\label{sect:core}
In this section, we consider the growth of symmetric shell-like clusters
around a spherical core particle A. The two cases that we consider are 
of an icosahedral and a dodecahedral outer shell, as depicted in 
Fig.\ \ref{fig:targets}. 
The B particles that make up the outer shell have $m$ identical patches whose 
positions are such that they point directly at the neighbouring B particles 
in the target geometry. These patches only interact with the equivalent 
patches on other B particles and with an interaction strength $\varepsilon_{\text{BB}}$. 
The B particles also have a second type of patch which points directly towards
the centre of the target clusters. These patches only interact with the A 
particles and have an interaction strength $\varepsilon_{\text{AB}}$. For this $\text{AB}$ 
interaction, there is no dependence on the orientation of particle A -- its contribution to the attraction is isotropic 
(or equivalently, $1/\sigma_{{\rm pw},A}=0$).

\begin{figure}
\includegraphics[width=7.4cm]{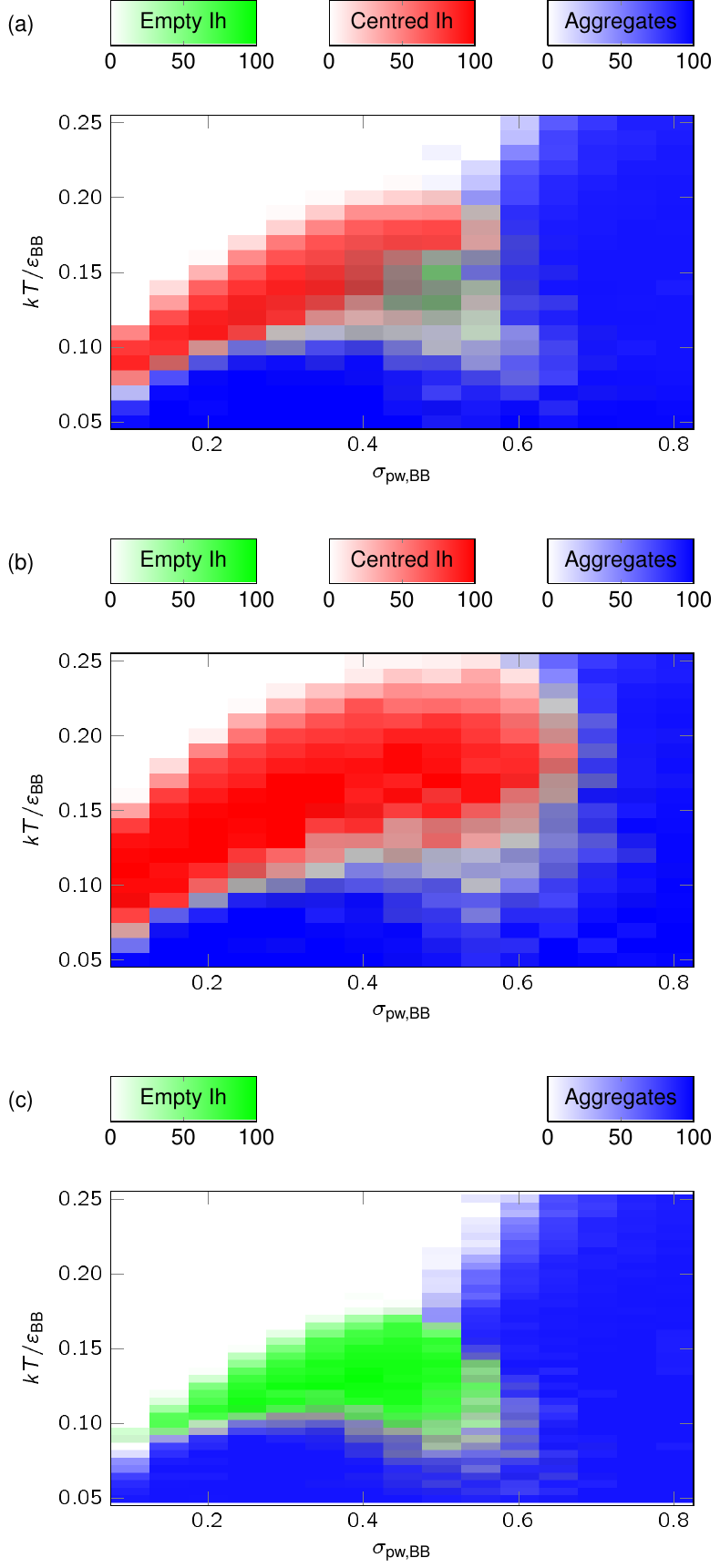}
\caption{\label{fig:Basic} (a) and (b) Yields (averaged over five repeats) of centred and empty icosahedra, 
and the number of particles in aggregates after $10^6$ MC cycles 
as a function of patch width (measured in radians) and the
temperature for a system of 120 B particles and 10 A particles at a density 
of B particles of $0.15\,\sigma_{\text{BB}}^{-3}$, where $\sigma_{\text{AA}}/\sigma_{\text{BB}}=0.95$
 and (a) $\varepsilon_{\text{AB}}/\varepsilon_{\text{BB}}=0.5$ and (b) 
$\varepsilon_{\text{AB}}/\varepsilon_{\text{BB}}=1.0$. 
For comparison in (c) we have plotted the yield of icosahedra and the number
of particles in aggregates under identical conditions but where only B 
particles are present.
Clusters are identified as centred and empty icosahedra if they have the correct number of particles and within two of the expected number of bonds, \emph{i.e.}\ 28--30 and 40--42 bonds for empty and centred icosahedra, respectively, where an interaction is considered a bond if it is at least 40\% of the well depth.  Aggregates are defined as clusters containing at least 25 particles.
} 
\end{figure}

Figure \ref{fig:Basic} shows the results for the assembly of centred icosahedra
as a function of patch width and temperature for two different values of 
$\varepsilon_{\text{AB}}$.
For comparison, the yield of icosahedra in simulations with only $B$ particles 
present is shown in Fig.\ \ref{fig:Basic}(c). In
this case, which we have considered in detail previously,\cite{Wilber07}  
there are two basic mechanisms of assembly. 
Firstly, at temperatures close to the clustering
temperature, $T_{\text{c}}$, at which icosahedral clusters become stable with respect to 
a gas of monomers, and at smaller $\sigma_{\rm pw}$, assembly of the icosahedra
proceeds by direct nucleation. By contrast, for wider patches and lower 
temperatures, large aggregates form first, but these can then undergo further 
internal rearrangements leading to the formation and budding off of 
complete icosahedra. However, at even lower temperatures, the time scale for the
rearrangements of the aggregates becomes so long that no icosahedra are able to
form, and at even wider patch widths liquid aggregates are thermodynamically stable with respect to icosahedra.

In our simulations with both A and B particles present, there is the 
potential for competition between the formation of centred and uncentred 
icosahedra. Interestingly, we find that 
for $\varepsilon_{\text{\text{AB}}}/\varepsilon_{\text{AA}}=0.5$ (Fig.\ \ref{fig:Basic}(a)), 
centred icosahedra are the dominant product in the region of parameter space 
that we previously identified as being dominated by direct nucleation for the 
pure B system, and that uncentred icosahedra preferentially form in the
region dominated by the budding-off mechanism. A closer comparison shows that
the centred icosahedra start to form at a somewhat higher temperature than for
the pure B system, which is unsurprising due to the extra stabilization due
to interactions with the central particle. In the region 
$T_{\text{c}}(\text{B}_{12})<T<T_{\text{c}}(\text{AB}_{12})$, there is no competition between the two forms as the only cluster stable 
with respect to the monomeric gas is the centred icosahedron and so growth is 
expected to occur by templated assembly around the central particle.

As one moves below $T_{\text{c}}(\text{B}_{12})$, the uncentred icosahedra can also start to 
form, and in the region where aggregation is initially more rapid than cluster 
formation, the uncentred icosahedra preferentially form. This preference arises because 
the formation of liquid aggregates is driven by the BB interactions and the A particles are 
generally excluded from the interior of these aggregates.
Again, at too low temperatures, the system gets trapped in aggregates, rather
than forming clusters.

\begin{figure}
\includegraphics[width=7.4cm]{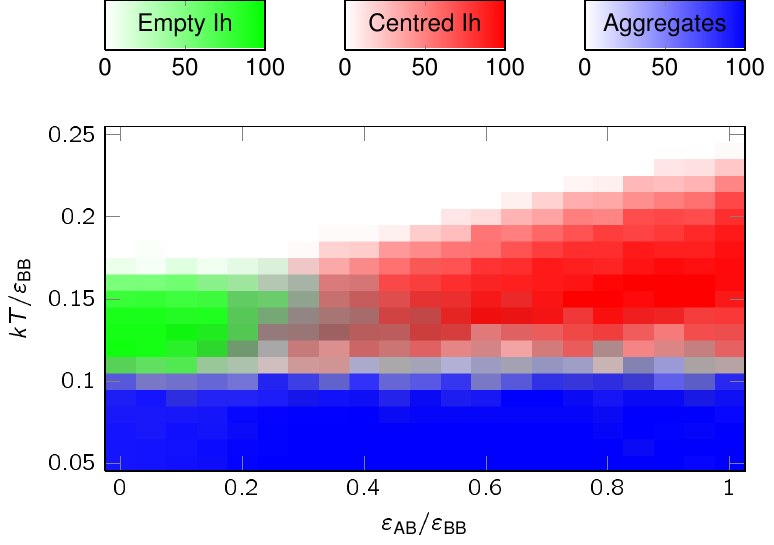}
\caption{\label{fig:Ih_epsilon} 
The yields of centred and empty icosahedra after $10^6$ MC cycles as a function of 
$\varepsilon_{\text{AB}}/\varepsilon_{\text{BB}}$ and temperature for a system of 120 B particles 
and 10 A particles at a density of B particles of $0.15\,\sigma_{\text{BB}}^{-3}$, 
where $\sigma_{pw}=0.35$ and $\sigma_{\text{AA}}/\sigma_{\text{\text{BB}}}=0.95$. Definitions of icosahedra and aggregates are the same as in Figure \ref{fig:Basic}.
}
\end{figure}

For $\varepsilon_{\text{AB}}/\varepsilon_{\text{BB}}=1$, (Fig.\ \ref{fig:Basic}(b)) the 
increased stabilization of the centred relative to the uncentred icosahedra 
is such that the centred icosahedra start to form at significantly higher temperatures than for the pure B system, and
 the uncentred icosahedra rarely form in any part of the parameter
space.

The effect of $\varepsilon_{\text{AB}}/\varepsilon_{\text{BB}}$ is explored further in 
Figure \ref{fig:Ih_epsilon}, where we look at the dependence of the competition 
between centred and uncentred icosahedra on $\varepsilon_{\text{AB}}/\varepsilon_{\text{BB}}$ at a
value of the patch width for which the yield of centred icosahedra
is high in Fig.\ \ref{fig:Basic} (a, b). 
At $\varepsilon_{\text{AB}}/\varepsilon_{\text{BB}}=1$, there is a wide range of temperature
over which centred icosahedra successfully assemble, and the
templated assembly of the icosahedron around A particles dominates over
direct nucleation of B$_{12}$ icosahedra. Below this temperature window,
the system forms kinetic aggregates (at this patch width, the local structure 
of these aggregate is not that similar to the target, so little product results from rearrangement of these aggregates, 
unlike at larger patch widths).


As $\varepsilon_{\text{AB}}/\varepsilon_{\text{BB}}$ decreases, $T_{\text{c}}(\text{AB}_{12})$ decreases, but the
temperature at which aggregation begins remains relatively unchanged, 
because this is mainly determined by $\varepsilon_{\text{BB}}$. Hence, the window over
which successful AB$_{12}$ assembly occurs decreases.
Furthermore, at sufficiently small $\varepsilon_{\text{AB}}/\varepsilon_{\text{BB}}$, $T_{\text{c}}(\text{AB}_{12})$ becomes lower than  
$T_{\text{c}}(B_{12})$ and at this point the energy gained from an A atom being inside
the B$_{12}$ icosahedron does not offset the loss of entropy. A simple 
estimate of the value of $\varepsilon_{\text{AB}}/\varepsilon_{\text{BB}}$ for this crossover 
can be found using the approximation that $T_{\text{c}}\propto E_{\text{gs}}/(n-1)$ 
where $E_{\text{gs}}$ is the ground state energy of the cluster and $n$ is the number
of atoms in the cluster; this expression has been found to provide a 
surprisingly accurate description of the dependence of transition temperatures
on relative patch strength.\cite{Villar09,Wilber09b,Villar10} 
Using $E_{\text{gs}}(B_{12})\approx -30\varepsilon_{\text{BB}}$ and 
$E_{\text{gs}}(\text{AB}_{12})\approx -30\varepsilon_{\text{BB}}-12\varepsilon_{\text{AB}}$ gives a crossover 
value of $\varepsilon_{\text{AB}}/\varepsilon_{\text{BB}}=5/22=0.227$.

However, even before this value of $\varepsilon_{\text{AB}}/\varepsilon_{\text{BB}}$ is reached,
the mechanism of assembly of AB$_{12}$ will have started to change. 
At $\varepsilon_{\text{AB}}=\varepsilon_{\text{AA}}$, 
$E(\text{AB}_{n-1})\le E(B_n)$ for any $n$ and so it is favourable for 
the $B$ particles to grow around the templating A particle. However, as
$\varepsilon_{\text{AB}}/\varepsilon_{\text{BB}}$ decreases, the value of $n$ at which it becomes
more energetically favourable to form AB$_{n-1}$ rather than B$_n$ increases,
and the role of A as a template diminishes. 
Consequently, at $\varepsilon_{\text{AB}}/\varepsilon_{\text{BB}}=0.227$, the direct nucleation of
uncentred icosahedra already dominates.


\begin{figure}
\includegraphics[width=7.4cm]{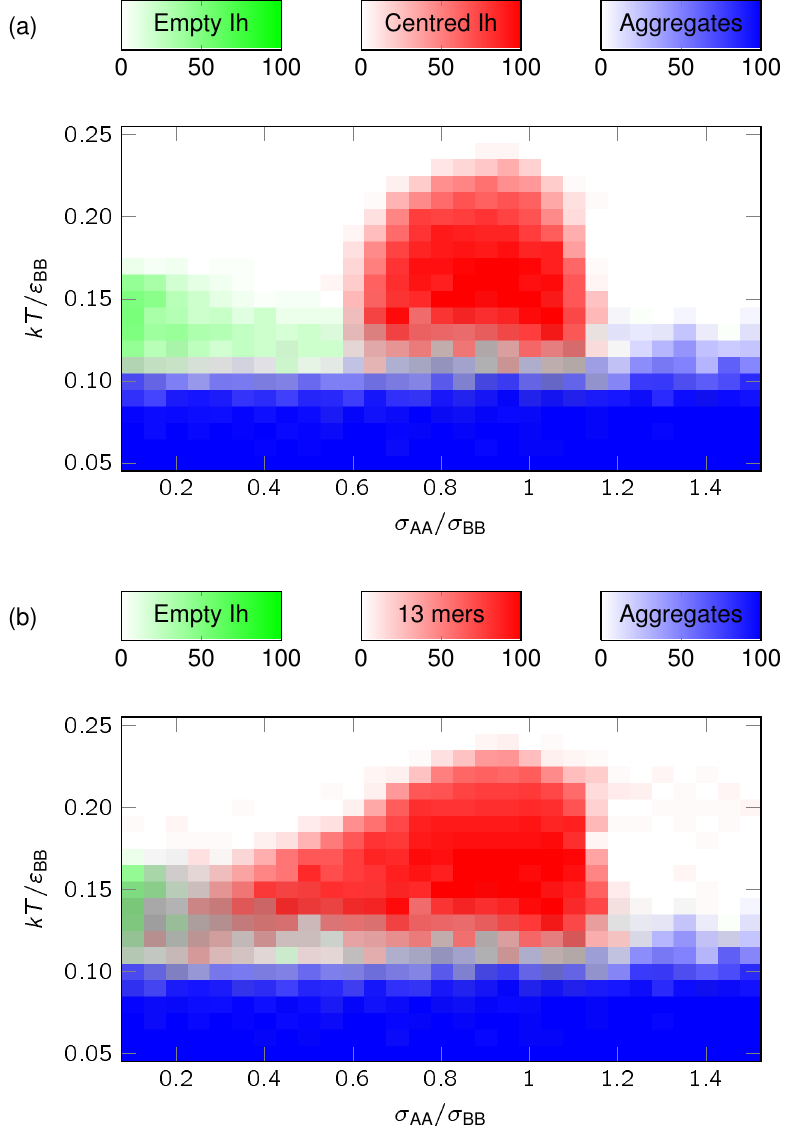}
\caption{\label{fig:Ih_size} 
The percentage yields of (a) centred icosahedra and (b) 13-mers.
$10^6$ MC cycles as a function of $\sigma_{\text{AA}}/\sigma_{\text{\text{BB}}}$ and 
temperature for a system of 120 B particles 
and 10 A particles at a density of B particles of $0.15\,\sigma_{\text{BB}}^{-3}$, 
where $\sigma_{pw}=0.35$ and $\varepsilon_{\text{AB}}/\varepsilon_{\text{BB}}$=1.  The definitions of icosahedra and aggregates are the same as in Figure \ref{fig:Basic}, and 13-mers are simply clusters with 13 particles.
}
\end{figure}

So far, we have considered cases where the A particle is the right
size to fit inside an icosahedron. In Figure \ref{fig:Ih_size}, we consider
the effect of the size of the A particle on the ease with which centred
icosahedra form. It can be seen that there is a limited size range over which
the centred icosahedra form. When $\sigma_{\text{AA}}/\sigma_{\text{\text{BB}}}$ is too large, 
although templated growth of $\text{AB}_n$ clusters will occur, the curvature of the
central particle is too small to allow the particles to form icosahedra. 
When $\sigma_{\text{AA}}/\sigma_{\text{\text{BB}}}$ is too small, although the initial nucleation
of the icosahedra still occurs on the A particle, not all the particles in 
the growing icosahedra can maintain contact with the A particle. Thus although AB$_{12}$ clusters still result, the A particle is now in an 
off-centre position touching only a subset of the B particles. 
Figures \ref{fig:Ih_size}(a) and (b) differentiate between the yields of 
AB$_{12}$ clusters with the A particle centred or off-centred. 
It is noticeable that the off-centred clusters persist to lower 
$\sigma_{\text{AA}}/\sigma_{\text{\text{BB}}}$ but that the temperature window over which they
can form decreases with decreasing $\sigma_{\text{AA}}/\sigma_{\text{BB}}$, because the
energetic stabilization of the cluster provided by the A particle decreases, 
as it can contact fewer and fewer of the B particles. Only at the smallest
values of $\sigma_{\text{AA}}/\sigma_{\text{BB}}$ do unfilled icosahedra begin to form.

Although icosahedral clusters can readily assemble in our simulations
with or without a templating central particle, the same is not true for
other target clusters. In particular, we previously found that it was 
impossible to get appropriately-designed patchy particles to form 20-particle 
dodecahedral shells.\cite{Wilber09} The essential problem is that the system 
always prefers to form aggregates rather than clusters, because the aggregates
 are first to become thermodynamically stable as the temperature is decreased, \emph{i.e.}\ $T_{\rm aggreg}>T_{\text{c}}(B_{20})$, 
and so there is no temperature window for which the target clusters
are the only species stable with respect to the gas. Furthermore, even when the dodecahedral clusters are more stable, \emph{i.e.}\ for
$T<T_{\text{c}}$ the aggregates form more rapidly than dodecahedral clusters.

Here we investigate whether we can get dodecahedral clusters to form using 
templated self-assembly, the idea being that the addition of templating 
particles could stabilize the dodecahedra sufficiently such that 
$T_{\text{c}}(\text{AB}_{20})>T_{\rm aggreg}$, thus resolving the thermodynamic problem noted
above.
Moreover, templated growth around
the central particles will help kinetically, by forcing the clusters to 
grow with the correct curvature. 

\begin{figure}
\includegraphics[width=7.4cm]{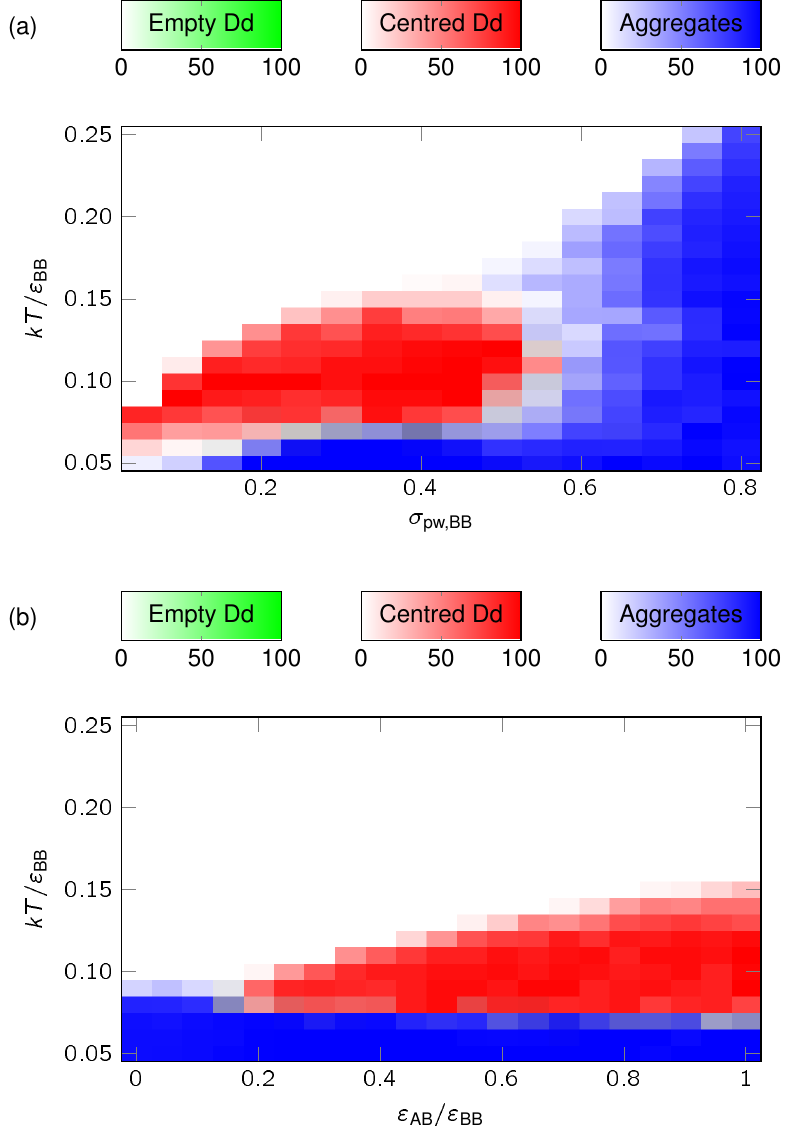}
\caption{\label{fig:dodecahedron} The yield of centred and empty dodecahedra
and the number of particle in aggregates (a) as a function of patch width
and temperature at $\varepsilon_{\text{AB}}/\varepsilon_{\text{BB}}=1.0$ and (b) as a function
of $\varepsilon_{\text{AB}}/\varepsilon_{\text{BB}}$ and temperature at $\sigma_{\text{pw}}=0.35$. 
The system consists of 10 A particles and 200 B particles at a density 
of B particles of $0.15\, \sigma_{\text{BB}}^{-3}$, where $\sigma_{\text{BB}}/\sigma_{\text{AA}}=1.80$. 
The simulations were of length $10^6$ MC cycles ($210\times 10^6$ MC steps).
Clusters are identified as centred and empty dodecahedra if they have the correct number of particles and within two of the expected number of bonds, \emph{i.e.}\ 28--30 and 48--50 bonds for empty and centred dodecahedra, respectively.  Aggregates are defined as clusters containing at least 41 particles.
}
\end{figure}

It can be seen from Figure \ref{fig:dodecahedron}(a) that there is now 
a clear region of parameter space where centred dodecahedra form.
 As expected, no uncentred dodecahedra form, and so the competition is simply between
templated assembly and aggregation.
The role of $\varepsilon_{\text{AB}}$ in stabilizing the target structure and enabling
assembly is clear from Figure \ref{fig:dodecahedron}, which shows the yield of 
$\text{AB}_{20}$ dodecahedra as a function of $\varepsilon_{\text{AB}}/\varepsilon_{\text{BB}}$. 
In the temperature window $T_{\rm aggreg}<T<T_{\text{c}}(\text{AB}_{20})$, dodecahedra now 
readily form, as the target clusters are the only species stable with respect to 
the gas.  Below this window, aggregation dominates because it occurs more 
rapidly than cluster formation, and because the aggregates are so structurally 
different from the target that rearrangement of the aggregates to form the 
target will never occur.

As $\varepsilon_{\text{AB}}/\varepsilon_{\text{BB}}$ decreases, there is a corresponding decrease
in $T_{\text{c}}$, and so the window of successful assembly first narrows and then 
disappears at $\varepsilon_{\text{AB}}/\varepsilon_{\text{BB}}=0.15$, 
where $T_{\text{c}}\approx T_{\rm aggreg}$. Below this value, similar
to the pure B system, dodecahedra are never found to assemble and 
aggregation always dominates.


\section{Multi-shell assembly}
\label{sect:multi}
The above results for the dodecahedra illustrate the potential role of 
templated self-assembly in enabling the assembly of more complex target
structures in systems of patchy particles without torsionally-specific 
interactions. In this section we take this one step further to consider 
the self-assembly of multi-shell structures, where the outer shell(s) surround 
a polyhedral core that we already know to assemble readily.

\begin{figure}
\begin{center}
\includegraphics[width=7.4cm]{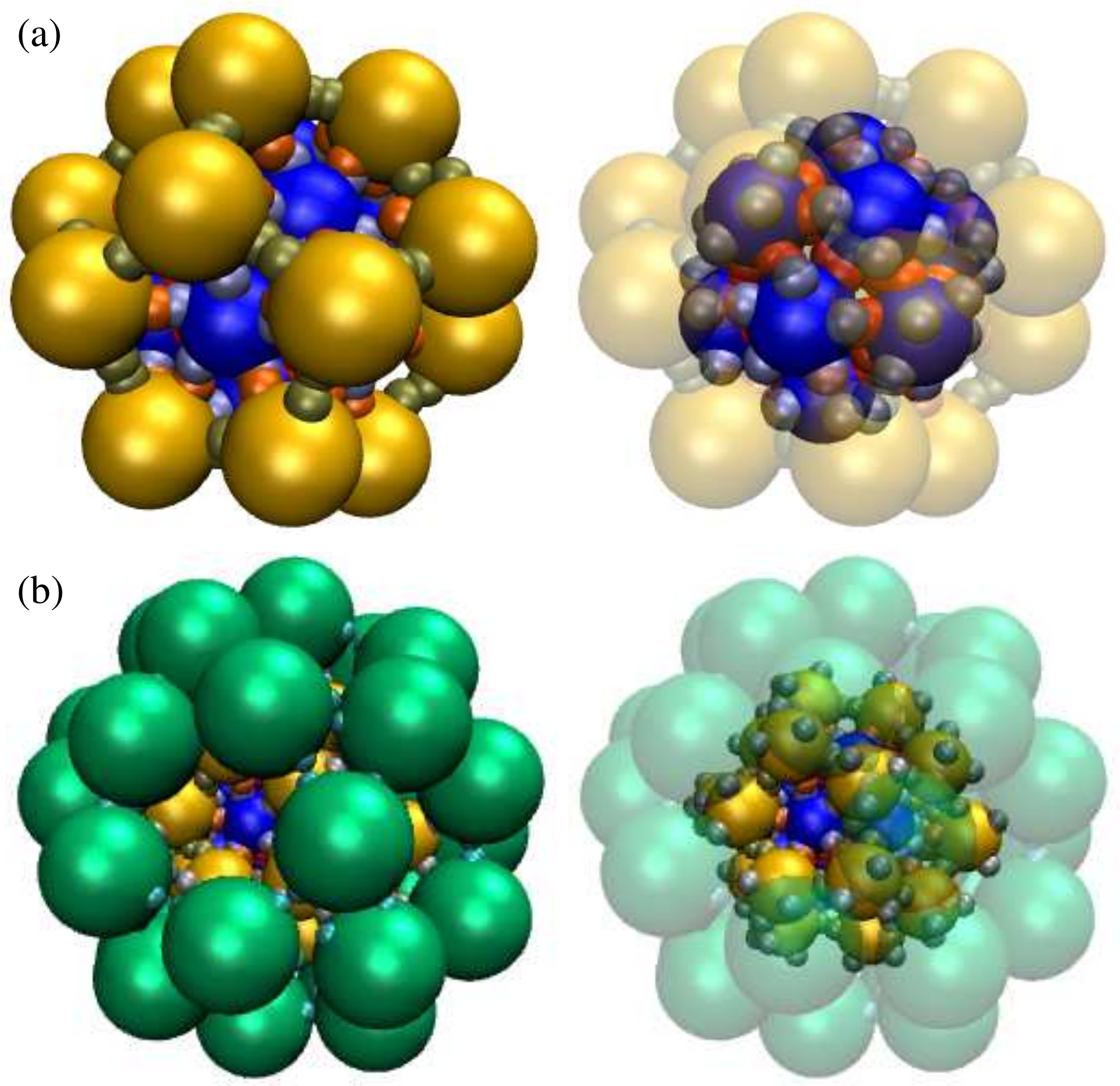}
\end{center}
\caption{\label{fig:multi-targets} Multi-shell target structures: 
(a) A$_{12}$B$_{20}$, a dodecahedron surrounding an icosahedron, and
(b) A$_{12}$B$_{20}$C$_{30}$, an icosidodecahedron surrounding a dodecahedron which itself 
surrounds an icosahedron.
There are two views of each structure, the ones on
the right having the outer shell partially transparent to give a clearer view 
of the core.  The clusters were produced by self-assembly at 
$T=0.14\, \varepsilon_{\text{AA}} k^{-1}$, and so show some deviations from the ideal
structure due to thermal vibrations.
The relative particle sizes were chosen to allow the structures to form without any strain,
i.e.\ $\sigma_{\text{BB}}=1.213\,\sigma_{\text{AA}}$ and $\sigma_{\text{CC}}=1.851\,\sigma_{\text{AA}}$.
}
\end{figure}

Similar highly-symmetric multi-shell structures are seen in biology. 
For example, some icosahedral viruses have capsids with two or more 
proteinaceous shells, either in their native state \cite{Chen09} or 
in malformed structures.\cite{Cardone09,Lavelle09} 
Furthermore, multi-enzyme complexes have been discovered that 
have an open multi-shell structure, e.g.\ in some species the pyruvate 
dehydrogenase complex has a dodecahedral inner core of 20 E1 trimers 
which can be surrounded by 60 E2 tetramers \cite{Milne03} or E3 
dimers\cite{Milne06} to give a complex with overall icosahedral symmetry.
However, the particles that we consider, unlike these proteins, do 
not have the advantage of torsionally-specific interactions.

The two example structures that we consider are illustrated in Fig.\ 
\ref{fig:multi-targets}. At the centre of both is an icosahedron of 12 
A particles. In the first example, this icosahedron is surrounded by a 
dodecahedron of 20 B particles. In the second example, in addition to the
dodecahedral shell, there is a further outer shell of 30 C particles 
with the structure of an icosidodecahedron. 

The general principle for choosing the geometry of the patches for these 
particle is that in the ideal target cluster there will be patches pointing 
directly at the neighbours both in the current shell and in the adjacent shell(s).
So, the A particles will have five patches that interact with strength 
$\varepsilon_{\text{AA}}$ with the equivalent patches on other A particles and that are 
responsible for forming the icosahedron. Each A particle will also have
a further 5 patches that point at the 
pentagon of B particles that surrounds each vertex of the icosahedron
in the target structure. These latter patches interact with strength 
$\varepsilon_{\text{AB}}$ with the three patches on the B particles that point to
the three A particles of the triangular face of the icosahedron that each B
particle sits above in the target cluster. Finally, the B particles also have a further
three patches that interact with strength 
$\varepsilon_{\text{BB}}$ with the equivalent patches on other B particles and that are 
responsible for forming the dodecahedron. Thus, the ground state energy of 
the A$_{12}$B$_{20}$ cluster is approximately 
$-30\varepsilon_{\text{AA}}-60\varepsilon_{\text{\text{AB}}}-30\varepsilon_{\text{BB}}$. 

For the A$_{12}$B$_{20}$C$_{30}$ cluster, the B particles have an additional 
three patches that point at the triangle of C particles that surrounds each 
vertex of the dodecahedron in the target structure. 
These latter patches interact with strength $\varepsilon_{\text{BC}}$ with the two 
patches on the C particles that point to the two B particles of the edge of 
the dodecahedron that each C particle sits above in the target cluster. 
The C particles also have a further
four patches that interact with strength $\varepsilon_{\text{CC}}$ with the 
equivalent patches on other C particles and that are 
responsible for forming the icosidodecahedron. 
Thus, the ground state energy of the A$_{12}$B$_{20}$C$_{30}$ cluster is approximately 
$-30\varepsilon_{\text{AA}}-60\varepsilon_{\text{AB}}-30\varepsilon_{\text{BB}}-60\varepsilon_{\text{BC}}-60\varepsilon_{\text{CC}}$.

The number of potential parameters when considering the assembly of these
multi-shell clusters is considerably larger than for the structures in Section 
\ref{sect:core}. Therefore, we always keep the size of the particles fixed
at their ideal values (as given in Fig.\ \ref{fig:multi-targets}), and for all 
patches $\sigma_{\rm pw}=0.4$. We always use a volume
fraction of 0.0785, which is equivalent to a number density of 
$0.15\,\sigma_{\text{AA}}^{-3}$ in a one-component system of A particles. 
We then consider the effects of varying the interaction strengths of the 
patches on the self-assembly behaviour.

\begin{figure*}
\centering
\includegraphics[width=16.5cm]{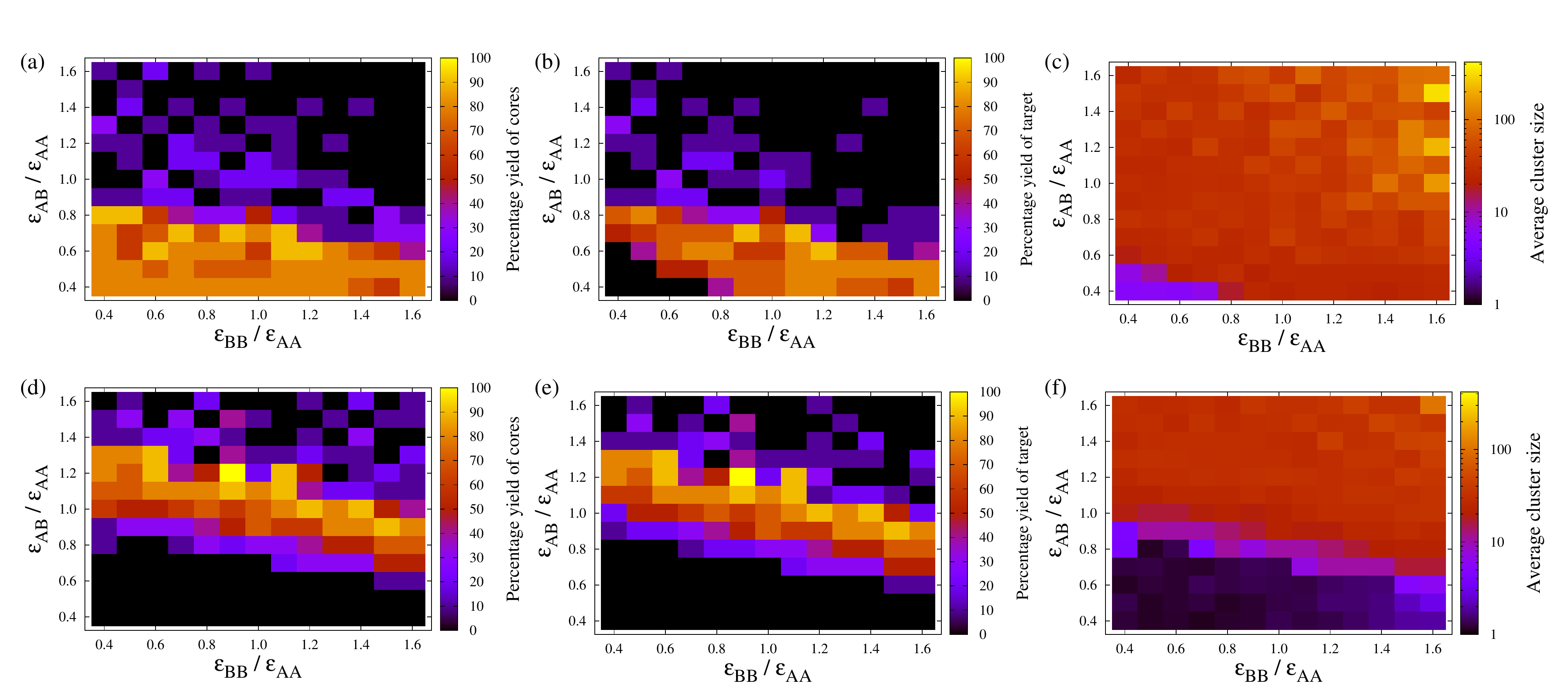}
\caption{\label{fig:ID} 
The dependence of the self-assembly of 
A$_{12}$B$_{20}$ on $\varepsilon_{\text{AB}}$ and $\varepsilon_{\text{BB}}$ at a given 
temperature: 
(a)--(c) $T=0.14\, \varepsilon_{\text{AA}} k^{-1}$, and
(d)--(f) $T=0.18\, \varepsilon_{\text{AA}} k^{-1}$. 
(a) and (d) show the yield of A$_{12}$ icosahedra (irrespective of whether 
they are bonded to B atoms),
(b) and (e) show the yield of A$_{12}$B$_{20}$, and 
(c) and (f) show the average cluster size, 
all after $625\,000$ MC cycles.
Each simulation contains 120 A and 200 B particles, so that a maximum
of ten target clusters could be formed. The volume fraction is 0.0785.}
\end{figure*}

Results for the self-assembly A$_{12}$B$_{20}$ are considered in 
Fig.\ \ref{fig:ID} as a function of $\varepsilon_{\text{AB}}$ and $\varepsilon_{\text{BB}}$
for two temperatures. 
At the first temperature we consider, 
$T=0.14\,\varepsilon_{\text{AA}} k^{-1}<T_{\text{c}}(A_{12})$, 
i.e.\ icosahedral A$_{12}$ clusters are stable irrespective of the values 
of $\varepsilon_{\text{AB}}$ and $\varepsilon_{\text{BB}}$. 
Thus, in the bottom left-hand corner of Figs.\ \ref{fig:ID}(a)-(c) 
corresponding to small $\varepsilon_{\text{AB}}$ and $\varepsilon_{\text{BB}}$, 
the formation of isolated A$_{12}$ clusters is observed. 
As either $\varepsilon_{\text{AB}}$ or $\varepsilon_{\text{BB}}$ is increased, 
A$_{12}$B$_{20}$ clusters become stabilized and a region of parameter space is 
reached where it becomes favourable to form these clusters. Close to
where they first become stable, A$_{12}$B$_{20}$ clusters readily form,
with yields of 80\% or more common. In this region, the majority of A$_{12}$ 
clusters are part of complete A$_{12}$B$_{20}$ clusters and provide a stable 
intermediate for the templated growth of the target cluster. 
The mechanism of assembly is likely to be hierarchical with A$_{12}$ clusters
forming first, followed by the growth of the second shell by the addition
of B monomers.

\begin{figure*}
\centering
\includegraphics[width=16.5cm]{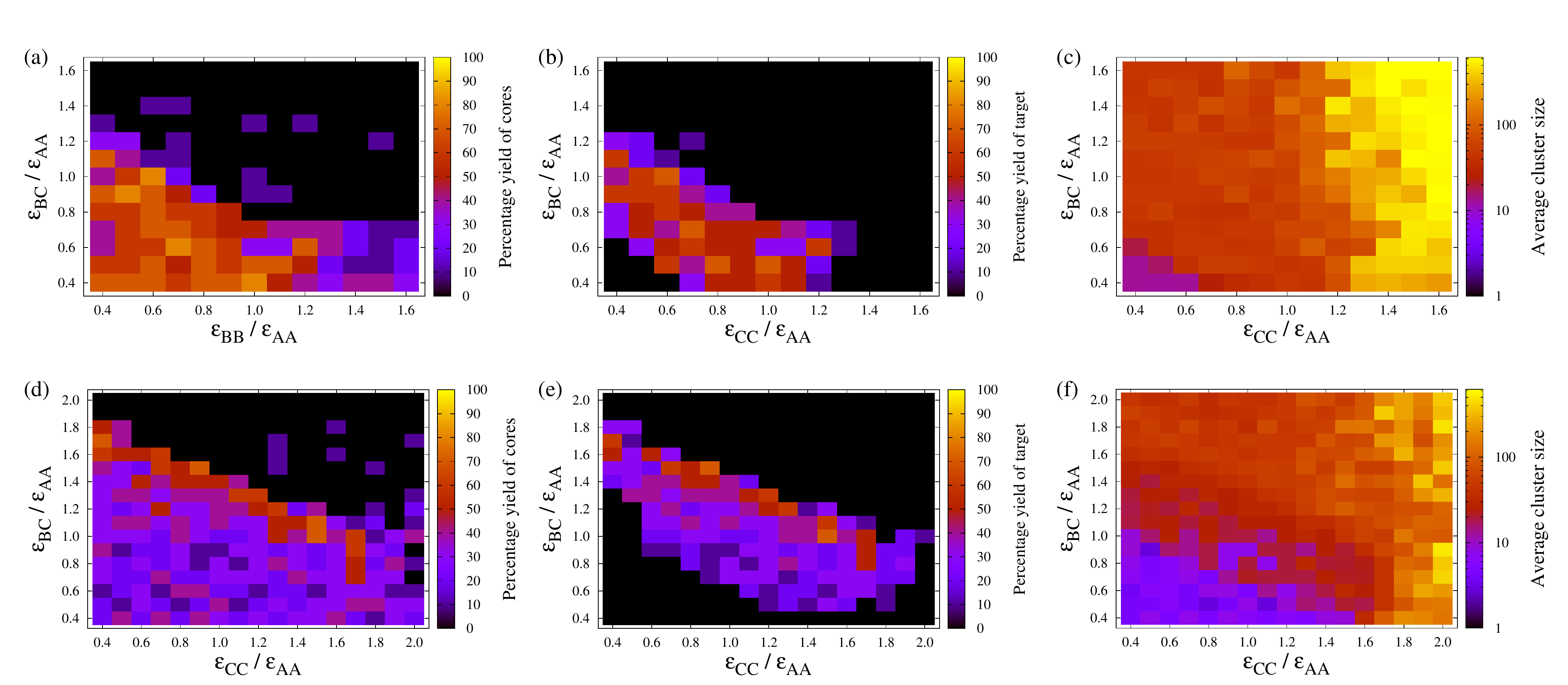}
\caption{\label{fig:IDC} 
The dependence of the self-assembly of 
A$_{12}$B$_{20}$C$_{30}$ on $\varepsilon_{\text{BC}}$ and $\varepsilon_{\text{CC}}$ at 
(a)-(c) $T=0.12\, \varepsilon_{\text{AA}} k^{-1}$, 
$\varepsilon_{\text{AB}}/\varepsilon_{\text{AA}}=0.5$ and $\varepsilon_{\text{BB}}/\varepsilon_{\text{AA}}=0.7$, and 
(d)-(f) $T=0.18\, \varepsilon_{\text{AA}} k^{-1}$, 
$\varepsilon_{\text{AB}}/\varepsilon_{\text{AA}}=1.1$ and $\varepsilon_{\text{BB}}/\varepsilon_{\text{AA}}=1.1$.
(a) and (d) show the yield of A$_{12}$B$_{20}$ (irrespective of whether 
they are bonded to C atoms),
(b) and (e) show the yield of A$_{12}$B$_{20}$C$_{30}$ and 
(c) and (f) show the average cluster size, 
all after $625\,000$ MC cycles.
Each simulation contains 120 A, 200 B and 300 C particles, so that a maximum
of ten target clusters could be formed. The volume fraction is 0.0785.}
\end{figure*}

However, as $\varepsilon_{\text{AB}}$, and to a lesser extent $\varepsilon_{\text{BB}}$, is 
increased further, malformed structures also become stable with respect to
the gas phase and the yield of both A$_{12}$ cores and 
A$_{12}$B$_{20}$ clusters falls off dramatically. It is noticeable that, unlike
the one-component self-assembling systems \cite{Wilber07,Wilber09}
and even the systems considered in
Sect.\ \ref{sect:core}, this fall-off in the yield is not 
associated with the formation of system-spanning 
aggregates, but instead the average cluster size remains similar
to that of the target cluster. Thus, there is a major difference 
in the configurations responsible for kinetic trapping
and this is because the trapping is dominated by interactions between 
different particle types rather than interactions between like particles. 
For example, we can see from Fig.\ \ref{fig:dodecahedron} that for aggregation between 
just B particles, $kT/\varepsilon_{\text{BB}}$ must be below approximately 0.07, 
which at $T=0.14\, \varepsilon_{\text{AA}} k^{-1}$ corresponds to 
$\varepsilon_{\text{BB}}/\varepsilon_{\text{AA}}>2$. 
Kinetic trapping is instead caused by the rapid binding between the A and B 
particles, before the A particles have assembled into icosahedra, due to the strength of the AB interactions, 
which in turn interferes with assembly for a number of reasons.
Firstly, there are kinetic effects. The B particles make it more difficult 
for the A particles to which they are bound to come together, both because 
they slow down the diffusion of the clusters and because the steric effect of 
the B particles means that it is less likely that a collision will lead to 
binding.  Furthermore, in collisions between clusters, these clusters would 
also have to have `matching' structures for them to be able to come together 
to generate a cluster that retains the correct structure (i.e.\ that is still 
a sub-cluster of the target structure). Secondly, there is the thermodynamic
problem that, due to the stabilizing effects of the AB interactions, clusters
with incorrectly formed structures for the A core are now stable with 
respect to the gas phase. Although the clusters formed have on average a similar
size to the target clusters, the system is typically a mixture of larger 
clusters with of the order of the 60 particles and smaller clusters with 
less than 10 particles.

At the second temperature that we consider, \mbox{$T=0.18\,\varepsilon k^{-1}$}, isolated
$A_{12}$ icosahedra are unstable, and so can only form due to the stabilization
provided by the second shell of B particles. Hence, in the bottom left-hand corner of Fig.\ \ref{fig:ID}(d) and (f) the system is now a mainly monomeric vapour.
Similar to the results for the lower temperature, as $\varepsilon_{\text{AB}}$ and $\varepsilon_{\text{BB}}$ are increased, there is a band in Fig.\ \ref{fig:ID}(e) associated 
with reliable self-assembly of the A$_{12}$B$_{20}$ target, before the yield
again falls off due to kinetic trapping. However, this band occurs at 
larger values of $\varepsilon_{\text{AB}}$ and $\varepsilon_{\text{BB}}$ as compared to Fig.\ 
\ref{fig:ID}(b) because stronger \text{AB} and BB interactions are required to 
compensate for the higher temperature. Furthermore, as A$_{12}$ clusters
are no longer a stable intermediate, the mechanism of assembly must 
be more cooperative. Indeed, it is noticeable from a comparison of Figs.\ 
\ref{fig:ID}(d) and (e) that virtually all A$_{12}$ clusters that form 
do so as part of the target structure.

These results show that the templating strategy is again successful in leading to
the formation of structures (dodecahedral shells) that are otherwise impossible 
to form. In the next example, A$_{12}$B$_{20}$C$_{30}$, we take this 
a stage further to show that templating can be used to create even more complex
structures. Given the large number of parameters for this system, we choose
$\varepsilon_{\text{AB}}$ and $\varepsilon_{\text{BB}}$ from within the region 
where A$_{12}$B$_{20}$ clusters were found to reliably form at the 
relevant temperature in the previous example. In Figure \ref{fig:IDC} we 
then show the self-assembly behaviour as a function of
$\varepsilon_{\text{BC}}$ and $\varepsilon_{\text{CC}}$ at two different temperatures.

\begin{figure}
\begin{center}
\includegraphics[width=6cm]{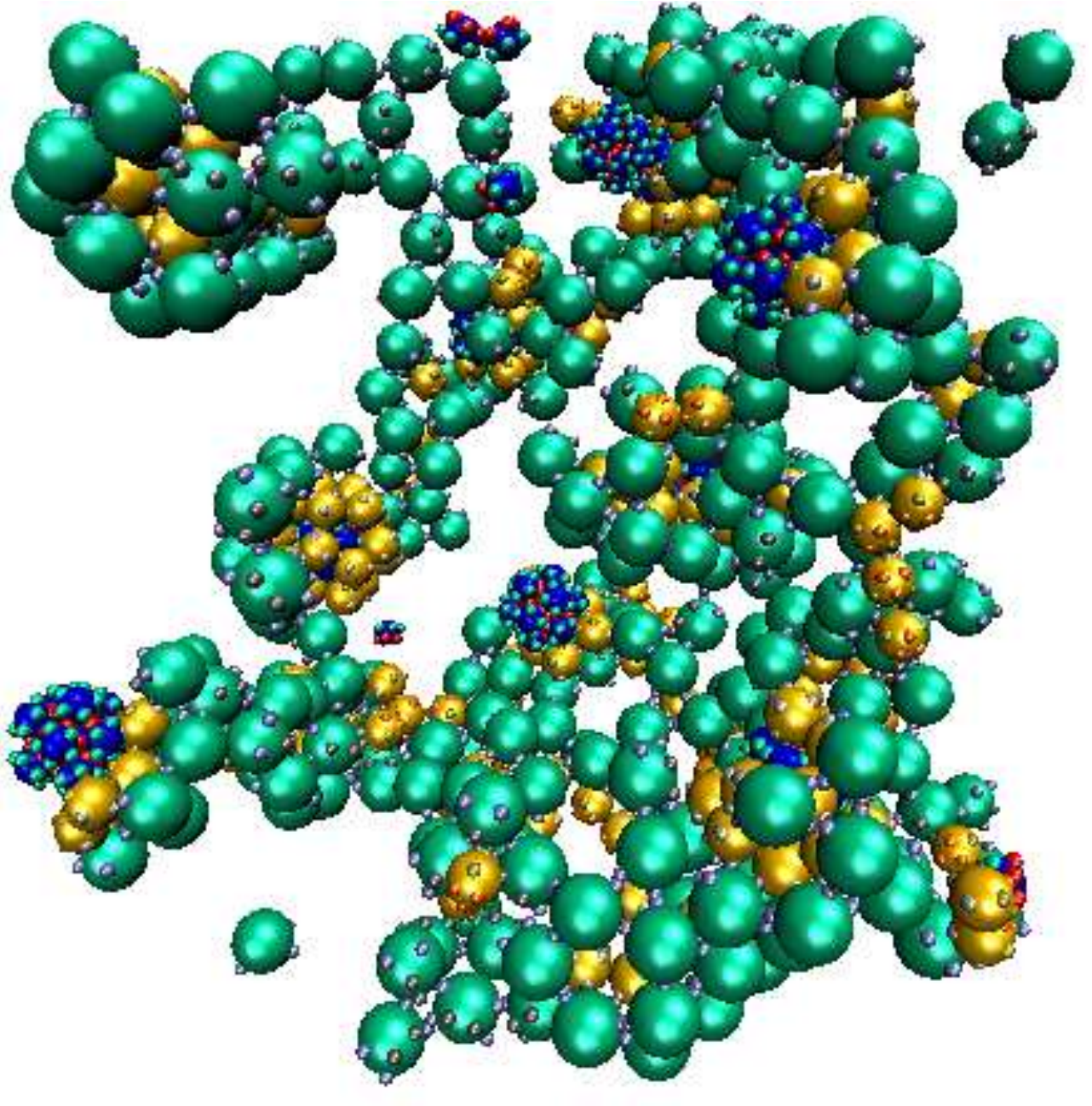}
\end{center}
\caption{\label{fig:IDCaggreg} Kinetic aggregates formed for the
A$_{12}$B$_{20}$C$_{30}$ target when 
$T=0.12\,\varepsilon_{\text{AA}} k^{-1}$, $\varepsilon_{\text{AB}}/\varepsilon_{\text{AA}}=0.5$, 
$\varepsilon_{\text{BB}}/\varepsilon_{\text{AA}}=0.7$,  
$\varepsilon_{\text{BC}}/\varepsilon_{\text{AA}}=1.2$,  and
$\varepsilon_{\text{CC}}/\varepsilon_{\text{AA}}=1.6$.}
\end{figure}

The self-assembly behaviour of this system is broadly similar to the 
that for the two-shell target. Again there is a diagonal band of successful
assembly as a function of the two interaction strengths 
(Fig.\ \ref{fig:IDC}(b) and (e)) with 
the interactions insufficient to stabilize the target 
at low $\varepsilon_{\text{BC}}$ and $\varepsilon_{\text{CC}}$, 
but so strong that the system becomes
kinetically trapped in incorrect configurations 
at high $\varepsilon_{\text{BC}}$ and $\varepsilon_{\text{CC}}$.

However, there are also a number of differences between the two systems. 
Firstly, the maximum yields are somewhat lower (about 70\%) but this is 
unsurprising given the greater complexity of the target. Perhaps
more surprising is that it decreases by so little; this is testament
to the robustness of the templated self-assembly approach. Secondly, there
is a stronger dependence of the behaviour on $\varepsilon_{\text{CC}}$ than there was on
$\varepsilon_{\text{BB}}$ in the previous example, but this is simply because there
are four such patches on each C particle (rather than the three for B 
particles). 
Consequently, $\varepsilon_{\text{CC}}$ plays a greater role in the stability of the
target structure (but also malformed structures). 
Thirdly, at large $\varepsilon_{\text{CC}}$ the formation of large aggregates is now
found to occur. Because of their four 
self-interactions, C particles 
can start to aggregate at higher values of $kT/\varepsilon_{\text{CC}}$. An 
aggregate from this region is illustrated in Fig.\ \ref{fig:IDCaggreg}.
Binding of B particles to the aggregates of C particles prevents the
formation of any A$_{12}$B$_{20}$ clusters, but A$_{12}$ icosahedra can 
still form because they have no interactions with the C particles.

There is also an interesting difference between the self-assembly behaviour of 
the systems at the two temperatures considered. 
Although for both temperatures $\varepsilon_{\text{AB}}$ and $\varepsilon_{\text{BB}}$ were
chosen with the aim of making A$_{12}$B$_{20}$ clusters stable, this seems
to have been only partially successful for the higher temperature 
with yields of only 30--40\% in the bottom left of Fig.\ \ref{fig:IDC}(d). 
Presumably, the lower yield is partly related to the fact that we have chosen 
to keep the overall volume fraction of particles the same in the two examples, 
meaning that there is a lower concentration of A and B particles compared to 
the two-shell case, and hence a lower driving force for A$_{12}$B$_{20}$ 
formation. Consequently, at $T=0.12\varepsilon_{\text{AA}} k^{-1}$, the assembly
mechanism can be hierarchical with A$_{12}$B$_{20}$ clusters a stable
intermediate, leading to a relatively broad band of high yield for the
target in Fig.\ \ref{fig:IDC}(b). By contrast, at 
$T=0.18\varepsilon_{\text{AA}} k^{-1}$ a more cooperative assembly mechanism is required
to achieve high yields, because of the lower stability of A$_{12}$B$_{20}$ 
clusters. In this case, as the interaction strengths increase 
there is initially a broad band of weak assembly of the target in Fig.\ \ref{fig:IDC}(e) with only the 30--40\% of particles that
are able to form A$_{12}$B$_{20}$ clusters going on to form the target. Only
at higher interaction strengths is there a narrow band of higher yield, 
presumably because the free energy barrier to direct nucleation of the 
A$_{12}$B$_{20}$C$_{30}$ target is now lower. 

\section{Conclusions}

In this paper, we have used computer simulations to investigate the efficacy of a templating strategy to facilitate the self-assembly of high symmetry monodisperse shell
structures from model patchy particles. 
Importantly, in contrast to the interactions between proteins, these particles lack a torsional component to the patch-patch interactions, thus providing a model for the synthetic patchy colloids and nanoparticles that many groups are seeking to develop, and allowing us to explore what structures it might be possible to assemble with such particles if the patch positions and the interaction strengths could be precisely controlled.
Previous work has shown that although simple target structures, \emph{e.g.}\ 12-particle icosahedra, can readily form, it is hard to form more complicated structures, because the lack of torsional specificity in the interactions means that the structure of the growing clusters is not tightly controlled. Consequently, there is typically a competition between correct assembly and the formation of disordered aggregates. 

The potential advantage of using a templating strategy is that it opens up a new assembly mechanism, namely heterogeneous nucleation around the template rather 
than direct homogeneous nucleation. The simple icosahedral example allowed us to explore under what conditions the templating pathway can dominate. 
Templating is generally more successful under conditions away from where aggregation offers a competing pathway, \emph{e.g.}\ higher temperatures and narrower patch widths.
Furthermore, the region dominated by templated assembly can be enhanced by increasing the interaction strength between the template and the assembling particles, and
thus opening up a larger temperature window over which the target structure is the only species stable with respect to the monomeric gas. 

We have then demonstrated the potential for templating to aid the formation of more complex structures by assembling dodecahedral and multi-shell clusters. 
The dodecahedral example is particularly noteworthy, because without the template, dodecahedra are never able to form as aggregation always dominates over assembly.\cite{Wilber09} 
However, the stabilization of the target structure by the template allows a temperature window to be opened up where the centred dodecahedra are the only species stable with respect to the monomeric gas, and in this region the dodecahedra can now assemble relatively easily. 
In addition to this thermodynamic effect, in removing aggregates as a competing state, 
templating also aids the dynamics of assembly by helping clusters to grow with the correct curvature and structure, even though their interactions are not torsionally specific. 
For example, although the angles between the patches allows dodecahedron-forming particles to form hexagonal as well as pentagonal rings, 
the former are likely to be disfavoured because they are less congruent with binding to the template.
Similarly, if as a cluster grows around a template, it incorporates some kind of defect, the propagation of this defect during further growth of the cluster is likely to inhibit the binding to the template and hence reduce the stability of the growing cluster, making the annealing out of that defect more likely.
Nevertheless, even given these advantages, it is impressive that templated assembly can allow us to form such complex structures as the three-shell A$_{12}$B$_{20}$C$_{30}$ cluster in relatively  high yields.

In our simulations, we are straightforwardly able to increase the number of particle types and types of patches, and to introduce specificity into the patch-patch interactions. By contrast, even though the synthetic strategies for producing patchy particles are rapidly improving, some of the particles whose behaviour we have analysed here would be very challenging to synthesise, particularly in terms of the control of patch position and identity---DNA-mediated interactions provide a potential route to achieve the required specificity in the patch-patch interactions.\cite{Maye09}  However, our results also indicate that the use of templated self-assembly provides a means to greatly increase the repertoire of structures into which such particles could assemble.

\section{Acknowledgements}
The authors are grateful to the EPSRC and the Royal Society for financial support.  We also thank Aleks Reinhardt for his expertise and support throughout this project.

\footnotesize{
\bibliography{references} 
\bibliographystyle{rsc} 
}

\end{document}